\documentclass[useAMS,usenatbib]{mn2e}
\usepackage{graphicx}
\usepackage{rotating}

\title[Hierarchical Star Formation]{Hierarchical Star Formation: Stars and Stellar Clusters in the  Gould Belt}
\author[F. Elias, E.J. Alfaro and J. Cabrera-Ca\~no]{F. Elias$^{1}$\thanks{E-mail:
felias@astroscu.unam.mx (FE); emilio@iaa.es (EJA)},  E.J.
Alfaro$^{2}$ and J. Cabrera-Ca\~no$^{3}$\\
$^{1}$Instituto de Astronom\'{\i}a. Universidad Nacional Aut\'onoma
de M\'exico. M\'exico, D.F., C.P. 04510\\
$^{2}$Instituto de Astrof\'{\i}sica de Andaluc\'{\i}a, CSIC, Apartado 3004 Granada, Spain.\\
$^{3}$Facultad de F\'{\i}sica. Departamento de F\'{\i}sica
At\'omica, Molecular y Nuclear. Universidad de Sevilla, Apartado
1065, Sevilla, Spain.}
\begin{document}

\date{Accepted 1988 December 15. Received 1988 December 14; in original form 1988 October 11}

\pagerange{\pageref{firstpage}--\pageref{lastpage}} \pubyear{2002}

\maketitle

\label{firstpage}

\begin{abstract}
We perform a study of the spatial and kinematical distribution of
  young open clusters in the solar neighborhood, discerning between
  bound clusters and transient stellar condensations within our
  sample. Then, we discriminate between Gould Belt (GB) and local
  Galactic disk (LGD) members, using a previous estimate
  \citep{Eli06a} of the structural parameters of both systems obtained
  from a sample of O-B6 \emph{Hipparcos} stars. Single membership
  probabilities of the clusters are also calculated in the separation
  process. Using this classified sample we analyze the spatial
  structure and the kinematic behavior of the cluster system in the
  GB. The two star formation regions that dominate and give the GB its
  characteristic inclined shape show a striking difference in their
  content of star clusters: while Ori OB1 is richly populated by open
  clusters, not a single one can be found within the boundaries of Sco
  OB2. This is mirrored in the velocity space, translating again into
  an abundance of clusters in the region of the kinematic space
  populated by the members of Ori OB1, and a marginal number of them
  associated to Sco OB2. We interpret all these differences by
  characterizing the Orion region as a cluster complex typically
  surrounded by a stellar halo, and the Sco-Cen region as an OB
  association in the outskirts of the complex. In the light of these
  results we study the nature of the GB with respect to the optical
  segment of the Orion Arm, and we propose that the different content
  of star clusters, the different heights over the Galactic plane and
  the different residual velocities of Ori OB1 and Sco OB2 can be
  explained in terms of their relative position to the density maximum
  of the Local Arm in the solar neighborhood. Although
  morphologically intriguing, the Gould Belt appears to be the result
  of our local and biased view of a larger star cluster complex
  in the Local Arm, that could be explained by
  the internal dynamics of the Galactic disk.
\end{abstract}


\section{Introduction}

The Gould Belt (GB) was first discovered by John \citet{Her47} and
Benjamin \citet{Gou79} as a system of bright stars inclined with
respect to the plane of the Milky Way. For more than a century, many
studies have been devoted to describing its structure and its
kinematical behavior, as well as to proposing a reliable global
scenario that would account for its origin (for an extensive review
on the subject, see P\"oppel 1997, 2001 and  Grenier 2004). Today it
is considered that, in the scope of the most recent theories of
hierarchical star formation \citep{Efr78,Efr95,Elm00,Elm06}, the GB
is likely to be our closest giant star forming complex
\citep{Com01}.

The stellar component of this complex takes the shape of a planar
distribution of bright and young OB stars inclined with respect to
the Galactic plane \citep{Les68,Sto74,Wes85}. Most of the young OB
associations in the solar neighborhood are known to be part of the
GB (Blaauw 1965; de Zeeuw et al. 1999; Elias et al. 2006a, hereafter
Paper I). Also, a system of young, low-mass stars, detected by
cross-matching X-ray and optical \emph{Hipparcos} \citep{Per97}
based catalogs, appears to be associated with the GB \citep{Gui98}.

As we should expect from a giant ongoing star forming complex, the
local interstellar medium is prominently associated with the GB. The
works by \citet{van66} on reflection nebulae, by \citet{San77} on
dark clouds and, more recently, by \citet{Gau93} on maps of infrared
emission, have found a spatial distribution of the dark clouds of
interstellar dust compatible with  the pattern shown by the stellar
component of the GB. \citet{Tom86,Tom87} and the study of the CO
molecule by \citet{Tay87} seem to confirm this inclination for the
local molecular clouds.

Neutral hydrogen in the solar neighborhood has also been related to
the GB after the discovery of Lindblad's ``feature A''
\citep{Lin67,Lin73}, interpreted as a ring of gas with an expansion
movement \citep{Ola82,Elm82}. After Lindblad's work, the giant
molecular clouds were found to be related to the most prominent OB
star associations \citep{Scs74,Kut77,Geu92}. A full map of the CO
molecule over the sky later confirmed that most of these clouds
within 1 kpc from the Sun follow the GB pattern \citep{Dam93}.

Then, if the GB is a star formation complex composed of both young
stars and associations and interstellar material, we would expect to
find a population of young open clusters following the GB pattern.
This is obvious if we consider the concept of young star cluster in
its widest sense, i.e., representing the dense inner regions of the
hierarchical structure of young star fields \citep{Elm06}.
Nonetheless, in this study we want to distinguish between a young
cluster understood as a mere stellar condensation, and a
gravitationally bound system (that is, as a condensation that ``has
had sufficient time and gravitational self-attraction to get mixed
by stellar orbital motions'', as put by Elmegreen 2006). Our work
focuses on the analysis of the cluster system, and on how these
objects distribute and behave in comparison with the GB defined by
the massive stars.

Since the first systematic studies that led to a discrimination
between the GB and the LGD \citep{Sto74,Wes85} there has been a
great improvement in the number of cataloged open clusters, as well
as in the availability of their astrometric and physical data.
Surprisingly, it is not possible to find in the scientific
literature any work devoted to the study of the open clusters
membership to the GB and their distribution within this system
before 2006. Thus, for many years, it has been accepted that the GB
did not contain a significant population of bound clusters. Only
very recently \citet{Pis06}, in their analysis of the
Galactic open cluster population, discovered an open cluster complex
(OCC) that they associate with the GB. Although they find this OCC
as a density peak within the spatial distribution of clusters in the
solar neighborhood, they estimate OCC membership probabilities by
kinematical methods, through the analysis of the tangential
velocities.

In two previous papers (Paper I; Elias et al. 2006b, hereafter Paper
II) we have studied the spatial distribution and the kinematic
properties of the OB stars and associations in the GB. Our line of
work thus leads in a logical way to the study of the distribution of
young open clusters in the GB and their kinematic properties. Our
analysis will be centered in the comparison of the GB morphology as
obtained from the distribution of massive stars and clusters. This
will represent another step in the understanding of the nature of
the GB, and will also contribute to the knowledge of how star
formation mechanisms proceed to the formation of stellar clusters.

\section{Open cluster system}

\subsection{Associations, stellar condensations and bound clusters}

Prior to describing the selection criteria for our observational
sample, we want to punctuate some ideas about the concepts of
hierarchical star formation, association, loose groupings and
clusters as gravitationally bound physical systems.

Stars are born from molecular gas clouds whose internal structure
can be characterized by a fractal dimension value that apparently
ranges from 2.5 to 2.7 (e.g. S\'anchez et al. 2005, 2007).
Simulations of the collapse of gas clouds hint at a hierarchical
structure of the stellar formation, with clusters present in several
dense cores \citep{Wal06}. During the past two decades, a set
of observations has been collected which indicates that young
stellar groupings show hierarchical patterns that range from the
larger scales of flocculent spiral arms and star complexes to the
smaller scales of OB associations, OB subgroups, small loose groups,
clusters and cluster subclumps \citep{Efr95}. In other words, the
newly born stars seem to follow the same fractal pattern as the gas
clouds from which they were originated.

The largest scale of stellar grouping, the great star complexes,
would be associated with the gas superclouds with masses the order
of 10$^{7}$ solar masses. According to some authors (e.g., Comer\'on
2001, Efremov \& Elmegreen 1998), the Gould Belt, with its spatial
scale the order of 1 kpc, would be the star complex closest to the
Sun, and thus would come from a single gas cloud with a mass of a
few million solar masses.

What happens at smaller scales? We already know that most of the OB
associations in the solar neighborhood, with typical sizes of about
80 pc, are mainly distributed along the plane of the GB. They
represent the observable spatial scale immediately below the star
complex. But, does it make any sense to
talk about a typical scale within a fractal structure? The answer
should be sought in the observational bias that is introduced by the
age limit of the sample: OB associations are detected and selected
as concentrations of OB stars with a typical age of about 10 Myr.
The existence of some general correlation between the duration of
the star formation and the size of the region \citep{Lar81,Elm06}
implies that, for a typical age of 10 Myr, the typical size of the
region is 80 pc \citep{Efr98,Elm06}.

Star clusters are formed in the cores of giant molecular clouds;
they represent the stellar groupings associated to the inner and
densest regions of the gas, and can be interpreted as the
unavoidable result of star formation in hierarchically structured
gas \citep{Elm06}. However, only a few of these condensations will
still be gravitationally bound after 10 Myr. It has been estimated
that 90\% of the clusters lose a high fraction of their stars in the
first 10 Myr of their lives \citep{Fal05}. Thus, after 10 Myr of
life, it is possible to find star forming regions that maintain a
large number of star clusters, while others like NGC 604 in M33 do
not contain a single cluster \citep{Mai01}. This seems to depend on
variations of the mean density of the clouds; those where the
average density is low will form stellar concentrations in the cores
of the clouds, but they will not have enough binding energy to keep
a bound cluster when the gas leaves. Thus, the clusters observed
within a star-forming region at a certain moment could be
representatives of two distinct physical systems: either star
clusters, gravitationally bound and able to survive galactic tidal
forces, or a mere stellar condensation with a mean lifetime the
order of 10 million years or less.

Thus the analysis of the distribution of star clusters (both bound
systems and transient condensations) could give us information about
the history of star formation in the GB, as well as about the
physical conditions of the gas from which they were born.

\subsection{Cluster sample}

We extract our sample from the Catalogue of Open Cluster Data (COCD)
and its Extension 1, compiled by \citet{Kar05a,Kar05b}. This
catalogue has the advantage of homogeneity over other existing
compilations, and since we are also interested in working with space
velocities, the fact that the COCD had catalogued proper motion
values in the \emph{Hipparcos} system and newly determined radial
velocity data was decisive in our choice.

As our aim is to study the cluster distribution in the GB, we
establish distance and age limits in our sample. The GB system
should be well contained within a heliocentric radius of 1 kpc
(Stothers \& Frogel 1974; Westin 1985; Paper I); and since its age
has been estimated between 20 and 90 Myr \citep{Tor00}, we only keep
clusters younger than 100 Myr.

Thus, our first selection (that will be reduced after outlier
elimination, as we explain in the following section) is composed of
93 open clusters, 83 of which have complete kinematical information.
We calculate for every cluster its Cartesian Galactic coordinates
($X$, $Y$, $Z$), where $X$ is positive in the direction of the
Galactic center, $Y$ in the direction of Galactic rotation and $Z$
perpendicular to the Galactic plane so that they form a
right-handed, orthogonal frame. We also calculate their respective
space velocities, ($U$, $V$, $W$), for those clusters with radial
velocity data.

\subsection{Gould Belt and Local Galactic Disk clusters}

In Paper I we developed a three-dimensional classification method
that allowed us to separate the GB stars from the local Galactic
disk (LGD) stars by purely spatial criteria. This method considered
that the LGD and the GB could be described as a distribution along
two intersecting planes; working with a sample of \emph{Hipparcos}
O-B6 stars we obtained an estimation of the parameters that defined
those planes. In Paper II, we used those parameters to classify
strictly by spatial criteria a new sample of OB stars with kinematic
data. Similarly, we now use our classification method and the
parameters found in Paper I to separate the GB open clusters from
those of the LGD field. We want to stress that this separation is
done only by means of their spatial position in the
three-dimensional space.

As described in Paper I, this classification method leads also to
the identification of outliers; i.e., those objects that are too far
away from the mean planes and thus lie in regions of low density of
probability in the sample space. We have found 9 outliers in our
cluster sample; after their elimination the remaining sample
consists in 84 clusters, 76 of them with full space velocity data
(listed in table~\ref{tbl-1}). Further information on the detection
and meaning of outliers may be found in Section 2 of Paper I and
references therein.

Finally, the individual GB membership probability ($P$) for each
cluster is listed in table~\ref{tbl-1}. In total, 40 clusters have a
probability $P > 50\%$, and thus, following the Bayes minimum error
rate decision rule, we consider them to be members of the GB.

\subsection{Bound and unbound clusters}

This sample may contain both bound and unbound clusters, as we have
previously seen (Section 2.1). A simple criterion to select those
condensations with a high probability of being bound clusters is an
age cut-off, if we consider that all the objects older than 10 Myr
can be called bound clusters, in the sense that they have survived
to the high rate of infant mortality that happens during the first
10 million years in the life of a cluster. Once this critical
threshold has been surpassed the mean life of a cluster, although
very dependent on the environmental conditions, is usually larger
than 1 Gyr.

It is evident, though, that such a selection is just an
approximation to the problem, and that to determine if a stellar
system represents a bound cluster, we must compare its density with
the tidal density at its position in the galaxy. However, we lack
the complete information to perform this detailed analysis, and thus
we must resort to empirical classification criteria. The COCD
catalog provides three variables for each cluster that may give us
some additional information about what we understand as young
stellar grouping and cluster; these variables are: cluster radius,
core radius and age.

In Figure 1, we represent the cluster sample in the cluster
radius-age space. It is interesting to note that the clusters older
than 10 Myr in our sample seem to group around an elongated strip
with a positive slope and that only a few objects present a cluster
radius that deviates from this mean distribution. The most notorious
case is that of NGC 2264, which has a radius close to 20 pc. This
cluster seems to present a high degree of substructure, spatial as
well as kinematical \citep{Fur06}, that its central potential has
not been able to erase. Thus we eliminate this object from the
sample, for its properties appear to differ in some structural
aspects from that of ``classic'' bound clusters.

We have also drawn in Figure 1 an upper envelope of the main cluster
distribution with ages older than 10 Myr and younger than 100 Myr.
There seems to be a natural separation between the main distribution
of clusters older than 10 Myr, and those that show a radius larger
than expected for their age. Based on this apparent separation, we
have traced this upper envelope of the main cluster distribution.
The exact location of the line has been drawn by consensus of the
authors. It simply aims to represent a dividing line between the
largest concentration of ``probable'' bound clusters (represented as
filled circles in the figure), and those that in spite of their age
show a clearly distinct behavior. The objects located under this
envelope line can be considered, from a conservative point of view,
as highly probable bound clusters.

Using the cluster radius and core radius data, we have estimated a
pseudo-concentration parameter in the form log(R$_{cl}$/R$_{co}$),
that we represent in Figure 2 versus the cluster radius. Circled
squares indicate those clusters that, according to the previously
exposed criteria, can be considered as probable bound clusters. We
note that most of these objects are distributed, as in Figure 1,
along a straight line. However, two of these objects (Platais 6 and
NGS 2546) show a clear separation from the general tendency.

Using the online tools at WEBDA database \citep{Mer95} we have
recalculated the parameters of these two clusters from their
photometric data. This analysis indicates that the photometric
diagrams of Platais 6 show a good fitting for a distance modulus of
7 and log(t)$\approx$ 6.5. Similarly, a visual inspection of the
color-magnitude diagram of NGC 2546, as cataloged by WEBDA, seems to
indicate that it is a very young cluster (log(t) $\le $ 7) with a
rich population of pre-main sequence stars. This diagram is similar
to that shown by NGC 2362 and other young clusters located in the
third Galactic quadrant \citep{Del06,Del07}, meaning that we are not
facing an object almost 100 Myr old. Thus we consider these two
objects as clusters with ages inferior to 10 Myr, and consequently
they do not belong --according to our criteria-- to the group of
clusters with a high probability of being gravitationally bound.

In table~\ref{tbl-1} we present the classification of the sample
that follows from this reasoning. The last column shows an indicator
of the probability of being a bound cluster, according to the
criteria previously discussed (1 stands for those objects that are
``probable" bound clusters, and 2 stands for those we consider as
transient stellar condensations).

\section{Analysis and discussion}

\subsection{Spatial distribution}

A two-dimensional projection of the spatial distribution of open
clusters in the sky is shown in Figure 3 (top panel). There we see,
as we commented in Section 1, that GB clusters (filled circles) are
mostly concentrated towards southern Galactic latitudes, and that
only three of them ($\sharp$395 = IC 4665, $\sharp$423 = Collinder
359, and $\sharp$456 = Stephenson 1) clearly rise above the Galactic
plane ($b > 10$ degrees). According to our criteria, $\sharp$395 and
$\sharp$456 are probable bound clusters, while $\sharp$423 could
rather be a transitory stellar condensation. We must note that these
three objects are located in the first Galactic quadrant and that,
apparently, are not related to any OB association.

We also represent in this figure the OB stars used in our analysis
of the spatial structure of the GB (Paper I). The associations Sco
OB2, Ori OB1, Per OB2 and Lac OB1 are classically thought to be
components of the GB (e.g Olano 1982), and studies that followed the
\emph{Hipparcos} mission suggest that Vel OB2, Tr 10 and Collinder
121 also belong to the GB \citep{Zee99}, although their position
close to the line of nodes where the GB intersects the Galactic disk
adds quite some uncertainty to this membership assignation. In any
case, it is evident that the Sco OB2 and Ori OB1 associations
respectively mark the South and North Galactic extremes of the
inclined plane which best describes the stellar system known as the
GB, and that in a certain sense the geometrical characterization of
the GB is defined by the relative position of these two
associations.

Thus, we have marked with a red cross the stars belonging to the Sco
OB2 association, following the coordinates as given by
\citet{Zee99}, that situate the complex in the range $l=290\degr -
360\degr$, $b=$ -10\degr $-$ 30\degr and D = 100 $-$ 220 pc, and
that we have tagged as Scorpius-Centaurus in the figure. In the same
manner, we have selected the stars belonging to Ori OB1 as those in
the range $l=197\degr - 215\degr$, $b=$ -12\degr $-$ -26\degr and D
= 300 $-$ 550 pc \citep{Zee99}. These stars appear in Figure 3 as
green triangles, and the region is tagged as Orion. Estimations of
the age range of these associations indicate that Sco OB2 members
are between 5 and 20 Myr old \citep{Sar03}, while the typical age of
Ori OB1 members is comprised between some 10$^5$ yr and 11 Myr
\citep{Bri05,Her06}.

Hence we are facing two associations with sizes and ages relatively
similar, and according to \citet{Zee99}, with a number of probable
Sco OB2 members larger than that of Ori OB1. With these data it
sounds reasonable enough to think that the star cluster population
related to Sco OB2 should be larger than that related to Ori OB1.
But what we observe in Figure 3A is exactly the opposite. There is
not a single cluster within the frontiers of Sco OB2, while we have
detected 11 objects associated to Ori OB1 ($\sharp$73, $\sharp$74,
$\sharp$75, $\sharp$76, $\sharp$77, $\sharp$80, $\sharp$1016,
$\sharp$1018, $\sharp$1019, $\sharp$1020, $\sharp$1021), six of them
being ``probable'' bound clusters ($\sharp$73=NGC
1981,$\sharp$74=NGC 1976, $\sharp$75=NGC 1977, $\sharp$1018,
$\sharp$1019, $\sharp$1020). This is better observed in Figure 3
(bottom panel), where we have represented only those clusters
cataloged as ``probable'' bound ones, tagged with number 1 in the
last column of table~\ref{tbl-1}. This figure shows how almost the
totality of the ``probable'' bound clusters associated to the GB are
located in the Orion and the Puppis-Vela regions. Moreover, almost
all of the other objects show no relationship with other OB
associations in the GB.

The same phenomenology can be observed in Figures 4 and 5, where we
have represented the GB cluster distribution in the spatial
projections $XY$ and $XZ$. The LGD clusters tend to distribute
uniformly across the Galactic plane in Figure 4, but that is not the
case for GB open clusters. We observe how the GB's distribution is
quite clumpy; many of the clusters are located in the region of
Vela, and we specially note that the region of Orion (in the third
quadrant of the $XY$ plane, and around the most negative values of
$Z$ in the Figure 5) presents an important concentration of members.
Moreover, if we consider only clusters with a higher membership
probability (75\%, as noted in Figure 6), and thus eliminating most
of the clusters around the line of nodes in which the GB plane
intersects the LGD plane, the GB is practically reduced to the Orion
region, as if that were the solely cluster population of the GB.

This analysis shows as an evident fact that while the Ori OB1
association is related with an important population of star clusters
(be they transitory stellar condensations or gravitationally bound
systems), the Sco OB2 complex does not include a single star cluster
within its frontiers, yet it is extremely rich in massive OB stars
and pre-main sequence stars \citep{Sar03}. In other
words, inside an apparently single star formation complex as the GB,
and separated only by about 500 pc, we find two star forming
regions, well defined both spatially and kinematically, that present
two clearly distinct modes of star formation. Orion is an OB
association that presents a large number of stellar condensations
seen as clusters, some of which appear to be gravitationally bound.
On the other hand, Sco-Cen seems to be forming only isolated stars
or loose groups that do not present the shape of a star cluster, and
thus have not been detected as such. But, how does this fact
translate to the velocity space?

\subsection{Kinematic behavior}

It seems obvious that the different spatial position of the
centroids of the two stellar groups analyzed in Section 3.1, implies
also a different location in the velocity space, specially in the
$V$ component due to their separation of almost 500 pc in the $X$
axis. The velocity centroids of these two associations, Ori OB1 and
Sco OB2, as defined by the GB stars, are located at (-16.4, -9.5,
-5.0) and (-8.0, -19.4, -6.0) km s$^{-1}$, respectively. This can be
clearly observed in Figure 7 (top panel), where we have represented
in the $UV$ plane the isodensity lines of the GB as defined by
massive stars, as well as the member stars of Ori OB1 and Sco OB2,
according to the criteria by \citet{Zee99} explained in Section 3.1,
and the GB star clusters. Once more, as it was expected, we observe
a clear correlation between the clusters' distribution and the Orion
velocity centroid, while the number of GB clusters associated to the
Sco-Cen complex is merely marginal.

The difference is even more evident if we limit ourselves to the
youngest clusters, which could be representatives of transient
stellar condensations (Figure 7, bottom panel). As we have commented
above, the difference observed between the velocity centroids was
predictable due to the differential Galactic rotation and to the
fact that both groups were quite separated in space. However, we
wonder if the rotation field in this region of the Galaxy can
completely explain the kinematic behavior of both the stars and the
clusters in the GB.

\citet{Mor99} analyzed the velocity space of a sample of OB stars
belonging to the GB, and found that the observed velocity field was
not compatible with that obtained from star formation models, as
well as with its dynamical evolution after the injection of momentum
and energy in the primeval cloud. In other words, the residual
velocities of the stars showed a highly negative vertex deviation
\citep{Fil57,Mih81} that could not be explained by the dynamical
model, because after a time interval quite inferior than the age of
the GB, the differential rotation prevailed over the movements
originated by the ``ad hoc'' energy and momentum injection, and gave
place to a slightly positive vertex deviation. \citet{Mor99} also
found that if the stars belonging to the Sco-Cen association were
eliminated, the vertex deviation became positive. That is, those
models designed to explain the origin of the GB from supernovae
explosions after a previous process of star formation, or from the
impact of a high velocity cloud on the Galactic disk (see P\"oppel
1997, 2001 and Sartori et al. 2003 for reviews about possibly
formation mechanisms of the GB) are not able to explain the observed
residual velocity field, unless the Sco-Cen stars are neglected.

Now we want to evaluate the residual velocity ellipsoid for the GB
clusters; in order to do so, we have corrected the velocities from
solar motion and differential rotation using the Oort constants: A =
16 km s$^{-1}$ kpc $^{-1}$ and B = -16 km s$^{-1}$ kpc $^{-1}$
(Paper II, M\'endez et al. 2000). These cluster residual velocities
are represented in Figure 8, along with the residual velocities of
OB stars, where different marks indicate those belonging to Ori OB1
and Sco OB2.  Then we calculate the longitude of the vertex, $l_v$,
for both systems, the LGD and the GB cluster samples. The result, if
we cut at GB membership probability of 50\%, is that $l_v = 9.3\degr
\pm 8.3\degr$ for the LGD, and that $l_v = -1.3\degr \pm 15.5\degr$
for the GB. This is a value very far from the GB vertex deviation of
$l_v = -47\degr$ found in Paper II for the OB stars belonging to the
GB. This is undoubtedly caused by the absence of clusters in the
Sco-Cen association, that was responsible for the large vertex
deviation of the GB (Moreno et al. 1999; Paper II). Moreover, if we
keep only GB clusters with a membership probability higher than
75\%, the vertex deviation of the system is $l_v = 9.7\degr \pm
16.3\degr$, which is practically the same as that of the LGD.

The vertex estimation for the star cluster system in the GB thus
gives us a double information. First, from a kinematic point of
view, the lack of star cluster associated to the Sco OB2 complex is
demonstrated. Second, the difference in the velocity space between
the Ori OB1 and Sco OB2 associations cannot be completely explained
by the Galactic differential rotation.

As we may see in Figure 8, the residual velocity distributions of
these two associations show different behavior. While the stars
belonging to Ori OB1 present a main velocity dispersion axis with a
positive vertex, the Sco OB2 stars present a main axis that is
almost perpendicular to the former, with a clearly negative vertex
deviation. Therefore, as it had already been noted (Moreno et al.
1999; Paper II), the vertex deviation in the solar neighborhood will
depend on the selection of the sample. If most of the sample stars
belong to Sco OB2, the vertex deviation will undoubtedly be
negative. If we extend our sample farther away from the Sun in order
to include Orion stars, we will reach some balance and thus the
vertex deviation will turn towards values closer to zero. The latter
is precisely what we observe in the star cluster population
associated to the GB: the lack of clusters within the Sco-Cen
complex makes the vertex deviation close to zero.

\subsection{Cluster complexes and scaled OB associations}

What we have observed when comparing the distribution of star
clusters and OB stars in the GB is that the two great complexes that
seem to define the North and South Galactic extremes of this large
stellar structure, show a different behavior according to the
scenery of hierarchical star formation. Ori OB1 shows a considerable
portion of its stellar population as grouped and forming star
clusters, half of them being probable bound clusters with ages
larger than 10 Myr. This kind of stellar system has been detected,
observed and analyzed in both the Milky Way \citep{EyS88,Alf91} and
other galaxies (e.g., 30 Doradus in the LMC; Hunter et al. 1995;
Walborn et al. 2002); not only it contains a rich star cluster
population, but also it is normally immersed inside a stellar halo.

On the opposite side of the star forming regions concentration
range, there can be found the scaled OB associations (SOBAs; e.g.,
Ma\'{\i}z-Apell\'aniz 2001), of which NGC 604 in M33 is a good
example. Although the star formation rate in NGC 604 is much higher
than that observed in the Sco OB2 association, they both have in
common their lack of star clusters. Thus, separated only by 500 pc,
there are two OB associations that, apparently, have been born from
molecular clouds under different ambient conditions, but that always
have been considered as the fundamental parts of a single stellar
system known as the GB. However, Ori OB1 seems to represent the
stellar halo associated to a cluster complex, while Sco OB2 appears
to be a clear example of an OB association, not related to cluster
formation.

If we interpret this result in terms of a hierarchical star
formation process \citep{Elm06,Elm08}, and considering that the age
of the stars in both associations present similar ranges, we should
consider either that the density maximum in the parental gas
distribution of Sco OB2 was inferior than the density peaks in the
clouds that formed Ori OB1, or that the tidal forces in the Sco-Cen
region were intense enough to destroy any substructure observable as
a star cluster, in an interval of time lesser than 10 Myr.

Any of these two possibilities requires some variations of the
ambient physical conditions in spatial scales smaller than 500 pc,
be they due to an external difference of pressure that caused higher
density peaks in the Orion region, or/and due to local gravitational
potentials or shear forces that shorten the lifetimes of the
transient stellar condensations in the Sco-Cen region.

\subsection{The Orion Arm}

Since the pioneering work of \citet{Bec56}, who traced the local
spiral structure from the young star clusters distribution within a
radius of 2 kpc, each time that a new catalog of star clusters has
been tailored, the corresponding map of this local spiral structure
has been drawn again (e.g., Janes \& Adler 1982). The results of
doing this show a series of cluster groupings with typical sizes the
order of 1 kpc \citep{EyS88,Alf91,Alf92} that seem to follow three
segments of spiral arms which have received the names of
Carina-Sagittarius Arm, Perseus Arm and Orion Arm (also known as
Local Arm). The inclusion of these optical segments within the
general scheme of the spiral structure of the Galaxy is
controversial, and depending on the spiral tracers and the analysis
techniques employed, different solutions for the number of arms,
their pitch angle or the velocity of the density pattern have been
found (see Naoz \& Shaviv 2007, and references therein).

In Figure 9 we represent the density map for the star clusters younger
than 10 Myr, inside a square of side length 4 kpc, centered in the
Sun. The data have been extracted from the COCD catalog. In the figure
we observe five main concentrations of young clusters that had
previously been detected by other authors (e.g., Efremov \& Sitnik
1988). In particular, the Orion and Cygnus complexes seem to align,
delimiting the local optical segment of the Orion Arm. Superimposed
over the young clusters density map, we have drawn the Ori OB1 and Sco
OB2 associations. The lines that depart from the centroids of both
associations represent their residual velocities, corrected from solar
motion and differential rotation.

Ori OB1 appears associated to the density maximum of the Orion Arm
defined by the young clusters; on the contrary, Sco OB2 is located
in the inner rim of the Arm, close to the Sun where the cluster
density is lower. If we consider that a higher density of clusters
is representative of a higher ambient pressure in the original gas,
the relative position of these associations in respect to the main
locus of the Orion spiral arm could explain their different content
of star clusters.

To better illustrate this, we define a Cluster Formation Index (CFI)
that describes the relative content of clusters with respect to the
OB stars that shape the GB:
\begin{equation}
CFI = \frac{Cluster~density}{Cluster~density + OB~star~density}
\end{equation}

The spatial densities of clusters and OB stars have been obtained by
using gaussian kernels in the GB plane ($X'Y'$), with a
$\sigma$~=~200 pc, and normalized so that the total sum of the
density be equal to 1. Then, Figure 10 shows the distribution in the
GB plane of the CFI parameter in the region that contains both OB
associations. A line that joins the centroids of both associations
has been represented over the density map and a cut along this line
(Figure 11), clearly shows a CFI gradient that ranges from 0.62 at
the maximum near Orion to 0.39 in the vicinity of Scorpius
Centaurus. Since the distance to the Ori OB1 centroid is a good
estimator of the distance to the Orion spiral arm, this indicates
that as we move away from the arm, the clustered star formation is
less efficient.

\subsection{The Nature of the Gould Belt}

The first detection of the GB was based in the fact that the
brightest stars in the sky, specially in the southern hemisphere,
present an eccentric position with respect to the great circle of
the Milky Way \citep{Her47,Gou79}. That is, it was a mere
morphological matter. Later studies determined the main stellar and
gaseous components of the GB, and from the analysis of their spatial
and kinematical properties, the shape, size and kinematics of the GB
were estimated.

Although the main parameters that describe the GB, published along
the past two decades, may differ depending on the authors (see Paper
I; P\"oppel 1997 and references therein), the general scheme of the
GB is similar for most of them: we are facing a star formation
complex with a disk-like structure and a radius of about 500 pc,
whose kinematic behavior is characterized by an expansion and a
rotation with respect to an internal axis. In most of the works that
contributed to the determination of these characteristics, the
separation between the probable members of the GB and the stars
belonging to the LGD has been performed. The analysis of the stellar
component of both groups has led to the conclusion that the GB and
the LGD are not only separated in the celestial sphere, but they
present different kinematic properties as well.

In most of these analyses it was necessary to perform a previous
separation of the elements belonging to either group, and in all
those cases such separation was mainly geometric (be it either in
the $l$-$b$ plane, in the different $XY$, $XZ$, $YZ$ projections, or
via a three-dimensional analysis). In other words, the stars that
are the brightest and the farthest from the Milky Way plane, and
that belong to the Sco-Cen and Orion constellations, seem to have
been the ones that opened the gates to this flood of studies, and
also the ones that, still today, best define and delimit the
geometry of the GB.

Then, we shall perform the following experiment: let us consider the
GB as defined only by two points whose coordinates in the phase
space are given by the spatial coordinates of the centroids of the
Ori OB1 and Sco OB2 association, and their velocities by their
respective central values of their residual velocities
(table~\ref{tbl-2}). A schema of the geometry of the problem is
shown in Figure 12. Using these values, we determine the inclination
($i$), the longitude of the ascending node ($\Omega$), the expansion
velocity along the line that joins the spatial centroids ($\rho_0$)
and the rotation velocity with respect to a point situated along the
main axis of the system ($\omega_0$), we reach the results listed on
table~\ref{tbl-3}, where we also present the range of values found
in the literature for the different variables in the GB.

These values are all very close to those of the fundamental
parameters of the GB that we can find in the literature (e.g., Paper
I; Fresneau et al. 1996; Lindblad 2000, among others). But on
sight of Figure 9, where we represent the position and the residual
velocities of both associations with respect to the Orion Arm, we
may ask ourselves if it makes any sense at all talking about a gas
and stars system with spatial and kinematic coherence that may be
described with some expansion and rotation velocities, or perhaps we
should definitely drop the traditional hypothesis of a single,
common origin for all the features of the GB, and begin to look at
it as a hazardous alignment -from out point of view- of at least two
of the many clumps in the hierarchical structure of the Local Arm,
with different densities and star formation history.
In this sense, the GB would be simply the projection over the sky of
the recent star formation in the clouds close to the Sun but located
far away from the fundamental Galactic plane. Therefore, it is
morphologically distinct, but does not necessarily represent a
physical system with unique properties, different from the star
formation within the Orion Arm.

We believe that the distinct properties of these two clumps --their
kinematic behavior, cluster content and height over the Galactic
plane- may be explained by the internal dynamics of the Galactic
disk. Further research should consider, as possible mechanisms
involved in the development of such a structure, the passing of a
density wave in a magnetized medium or the presence of a long bar
generating resonances in the external parts of the disk
\citep{Gar08}.

Thus, if we could observe our Galaxy from an external position
several megaparsecs away, what would stand out in the solar
neighborhood? Undoubtedly, the complex of blue star clusters inside
the Local Arm, where the Ori OB1 and Sco OB2 associations, that form
the main structure of the GB, would just be accessory elements
related to the nucleus of the complex and its periphery,
respectively.

\section{Conclusions}

From a sample of star clusters younger than 100 Myr, and located
within a radius of 1 kpc around the Sun, we have analyzed the
spatial and kinematical structure of the GB. The comparative
analysis between the populations of stellar clusters and OB stars in
the GB, indicates that the Ori OB1 and Sco OB2 associations present
a significant difference in the number of clusters related to them,
and that this difference is even more conspicuous in the case of the
transient stellar condensations. While Ori OB1 can be characterized
as the stellar population associated with the core of a star cluster
complex (e.g., 30 Doradus), the Sco-Cen complex apparently shows a
star formation mode where the generation of isolated stars is
dominant.

In the light of this scenario of hierarchical star formation, this
difference in the content of stellar clusters must have been caused
by different physical conditions of the primeval clouds and/or
inhomogeneities of the gravitational potential.
The main physical characteristics of these two large associations
are:

\begin{enumerate}

\item Different height over the Galactic plane
\item Different content of stellar clusters
\item Different residual velocity vectors

\end{enumerate}

\noindent All these differences can be explained, at least
qualitatively, by the different position of these two associations
with respect to the main loci of the young stellar clusters that
define the Orion Arm. According to this scenario, the GB can be
considered as a partial and biased vision of a much larger scale
process of stellar formation, which is currently visible as a star
cluster complex in the region of the Orion Arm that is closest to
us.

\section*{Acknowledgments}

We thank the referee for the useful comments. This research has made
use of the WEBDA database, operated at the Institute for Astronomy
of the University of Vienna, and the NASA's Astrophysics Data
System. We would like to acknowledge the funding from MICCIN of
Spain through grant AYA2007-64052 and from Consejer\'{\i}a de
Innovaci\'on, Ciencia y Empresa (Junta de Andaluc\'{\i}a) through
TIC-101.  This work has been partially funded by the
CONSOLIDER-INGENIO-2010 program from MEC (Spain) through project
{\sl First Science with the GTC}.

\clearpage
\begin{figure}
\begin{center}
\includegraphics[angle=270,scale=0.60]{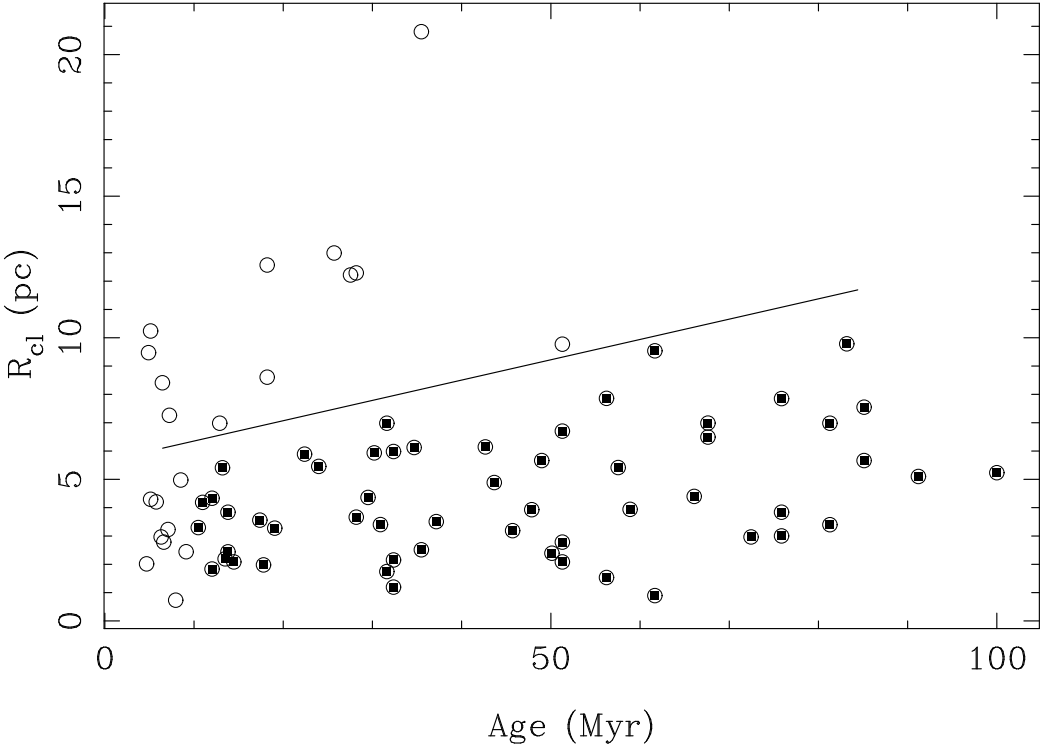}
\caption{Cluster radius vs. age. The filled circles represent {\sl
our probable} bound clusters in the age range from 10 to 100 Myr.}
\end{center}
\end{figure}
\clearpage

\clearpage
\begin{figure}
\begin{center}
\includegraphics[angle=270,scale=0.60]{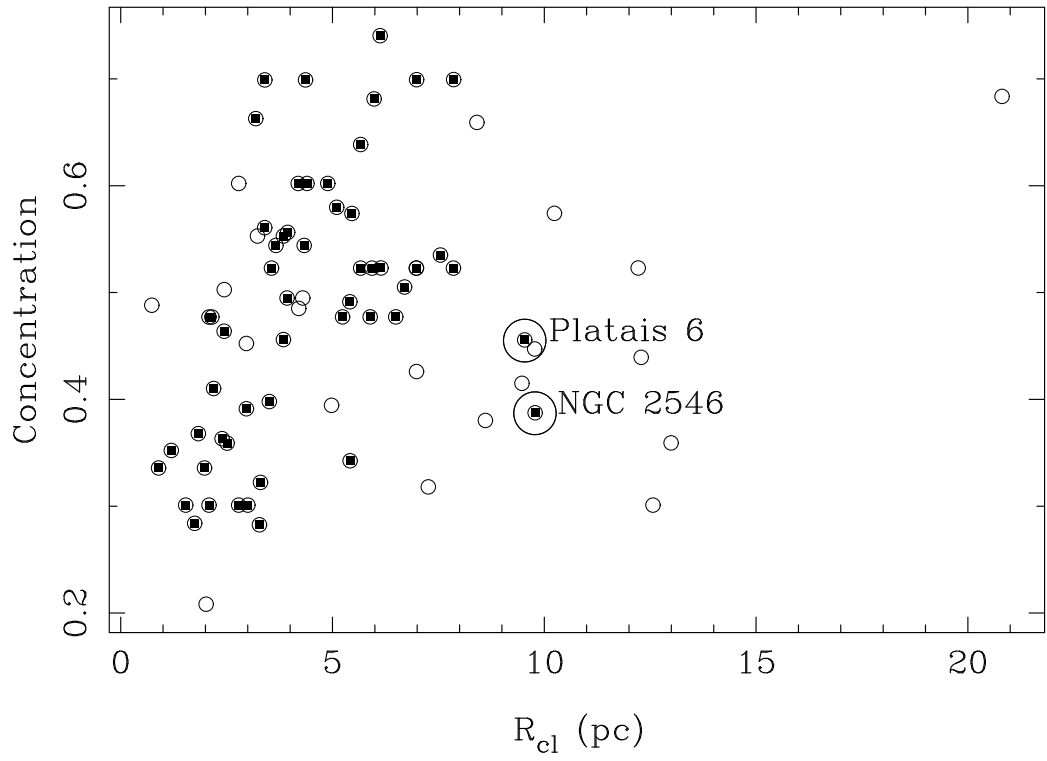}
\caption{Pseudo concentration parameter, log(R$_{cl}$/R$_{co}$) vs.
cluster radius. Filled circles represent {\sl our probable}  bound
clusters.}
\end{center}
\end{figure}
\clearpage

\clearpage
\begin{figure}
\begin{center}
\includegraphics [angle=0,scale=0.80]{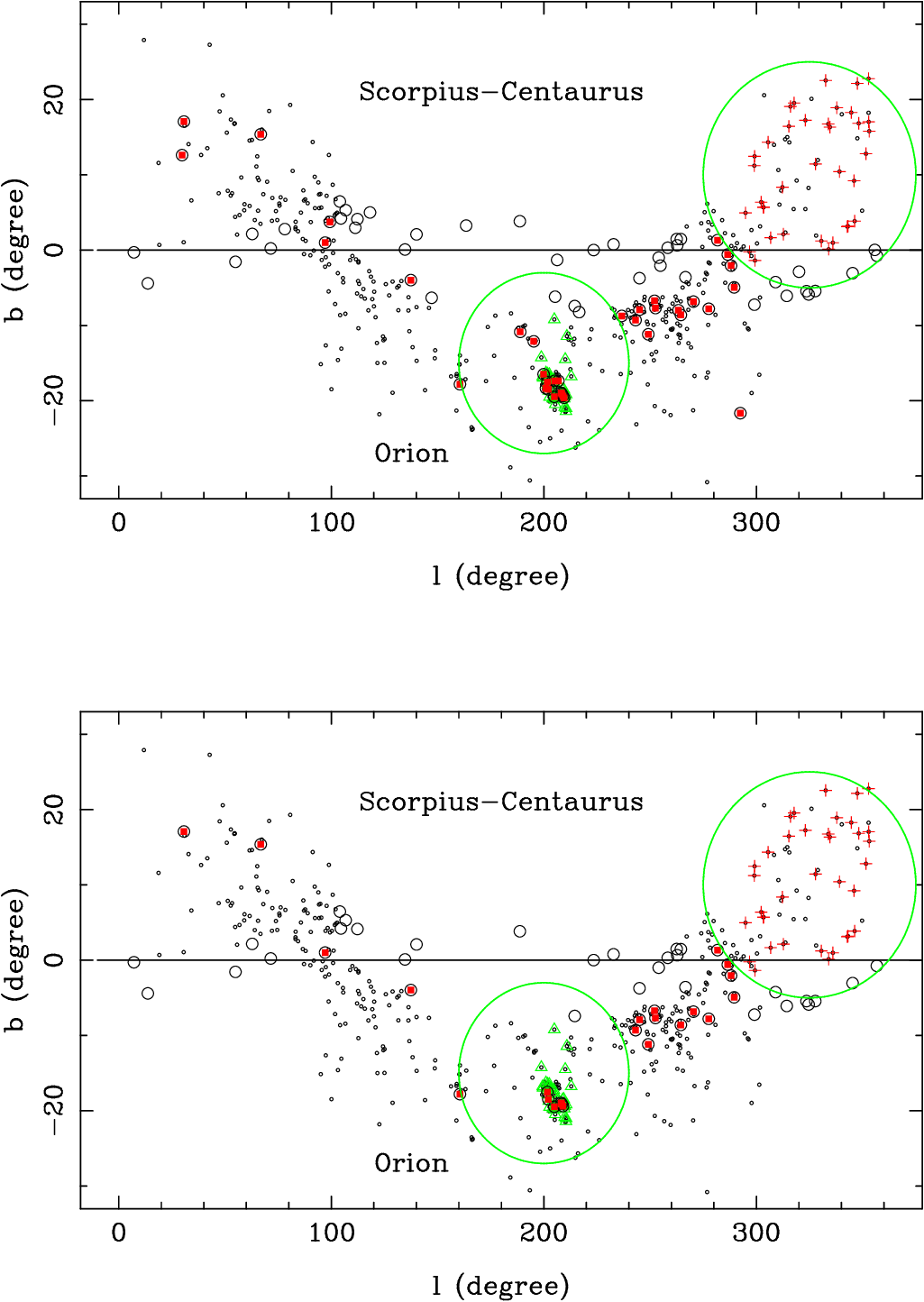}
\caption{Distribution of open clusters on the sky (galactic
coordinates). Open and filled circles represent clusters classified
as LGD and GB members, respectively. Small circles stand for the O-B6
GB stars from Paper I. Red crosses and green triangles mark,
respectively, the stars belonging to Sco OB2 and Ori OB1, according
to \citet{Zee99}. Top panel shows all the clusters in our sample,
whereas bottom panel displays only the probable bound clusters.}
\end{center}
\end{figure}
\clearpage

\clearpage
\begin{figure}
\begin{center}
\includegraphics[angle=270,scale=0.60]{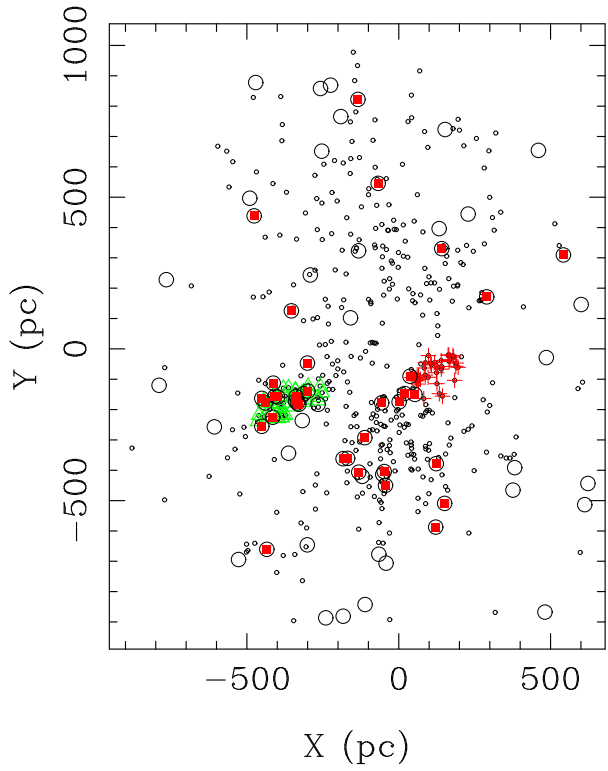}
\caption{Spatial distribution of the sample clusters in the $XY$
plane. Symbols as in Figure 3.}
\end{center}
\end{figure}
\clearpage

\clearpage
\begin{figure}
\begin{center}
\includegraphics[angle=270,scale=0.60]{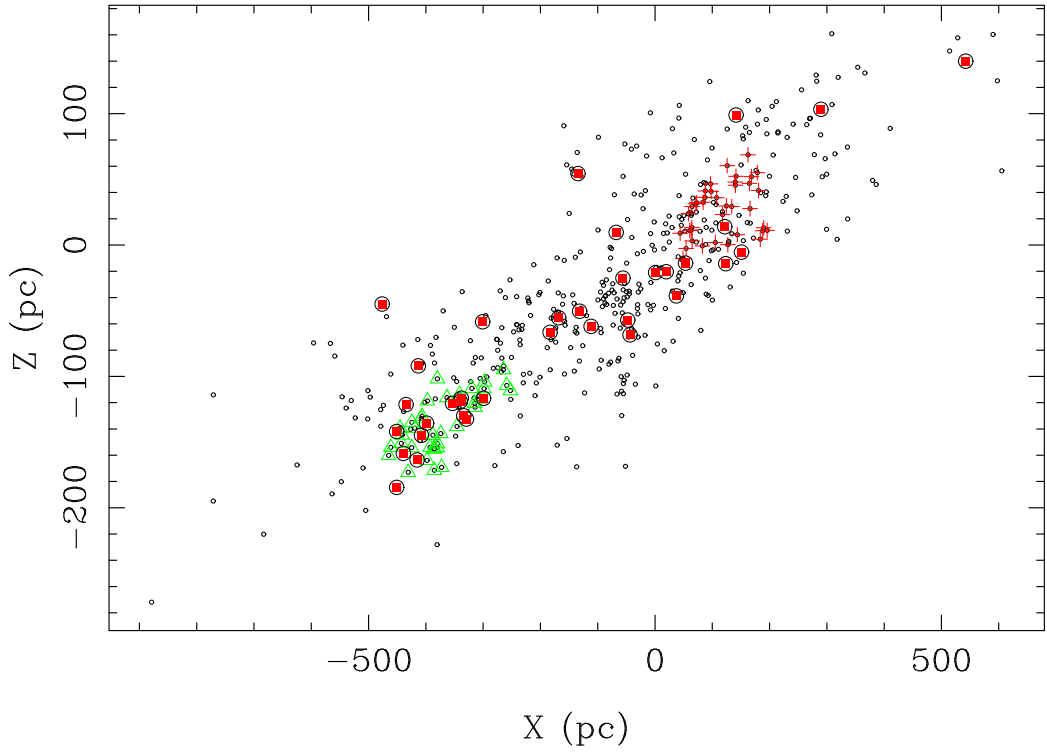}
\caption{Spatial distribution of GB clusters (filled circles) in the
$XZ$ plane. Other symbols as in Figure 3.}
\end{center}
\end{figure}
\clearpage

\clearpage
\begin{figure}
\begin{center}
\includegraphics[angle=270,scale=0.60]{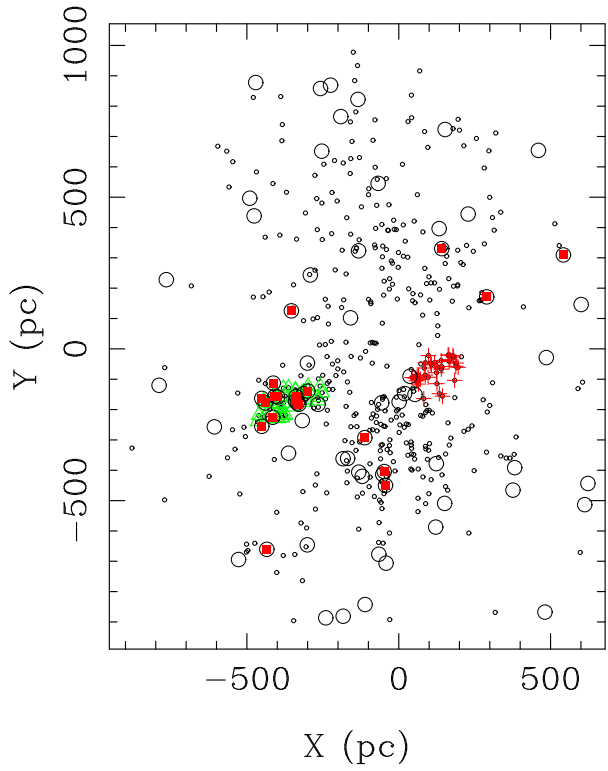}
\caption{Same as Figure 4, but including only GB clusters (filled
circles) with a membership probability higher than 75\%. Other
symbols as in Figure 3.}
\end{center}
\end{figure}
\clearpage

\clearpage
\begin{figure}
\begin{center}
\includegraphics[angle=0,scale=0.80]{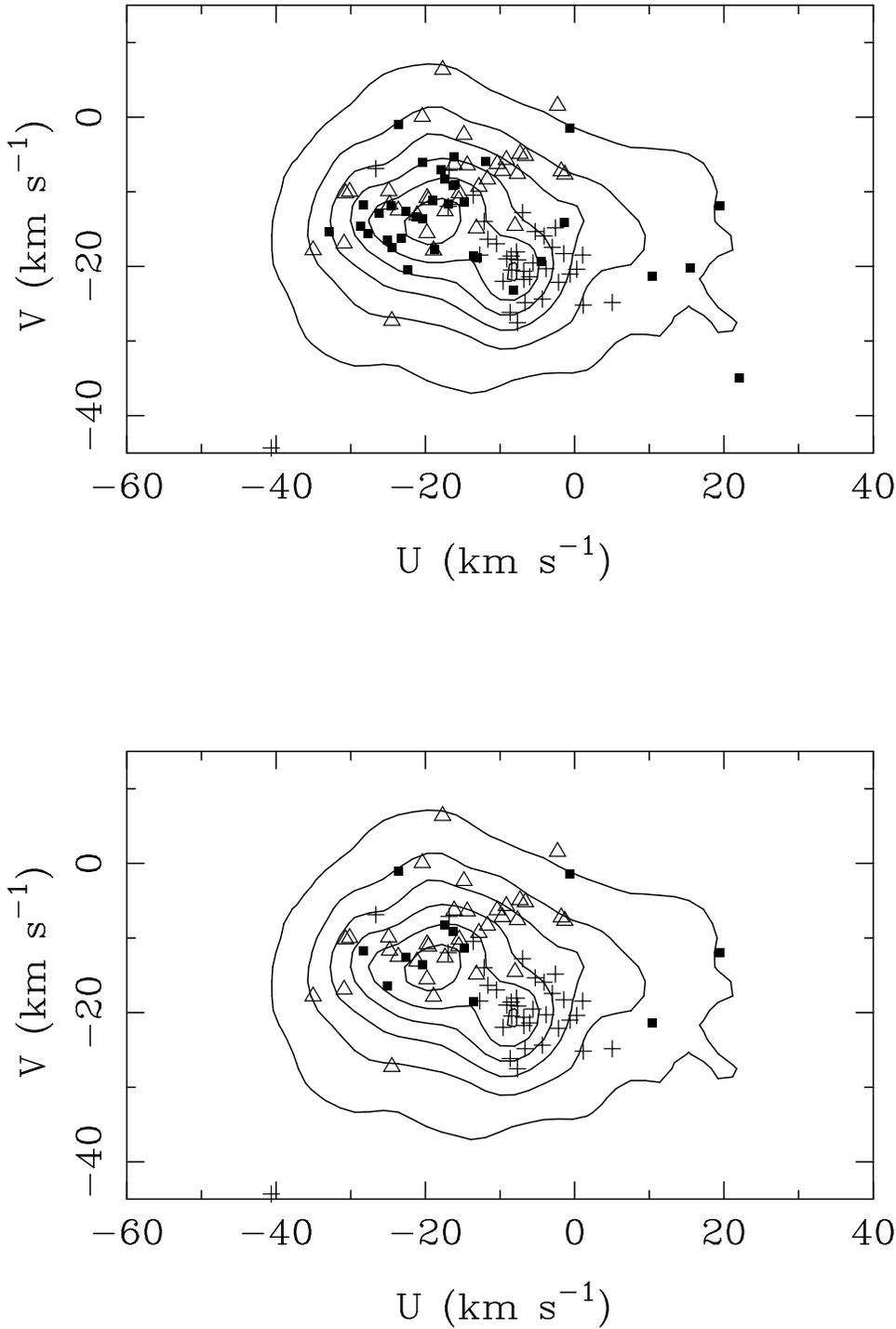}
\caption{Velocity distribution of GB clusters (filled squares)
against that of the GB stars (isodensity lines) from Paper I.
Crosses and open triangles mark, respectively, the stars belonging
to Sco OB2 and Ori OB1, according to \citet{Zee99}. Top panel
represents the totality of GB clusters, whereas bottom panel
displays only those classified as {\sl probable} transient stellar
condensations.}
\end{center}
\end{figure}
\clearpage

\clearpage
\begin{figure}
\begin{center}
\includegraphics[angle=270,scale=0.60]{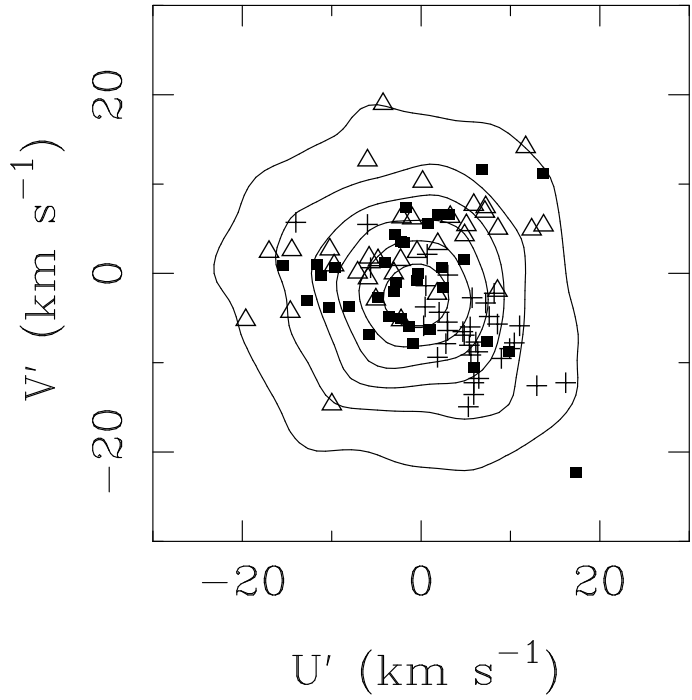}
\caption{Distribution of the residual velocities, corrected from
solar motion and differential rotation,  of GB clusters (filled
squares) against that of the GB stars (isodensity lines) from Paper
I. Crosses and open triangles mark, respectively, the stars
belonging to Sco OB2 and Ori OB1, according to \citet{Zee99}.}
\end{center}
\end{figure}
\clearpage

\clearpage
\begin{figure}
\begin{center}
\includegraphics[angle=270,scale=0.60]{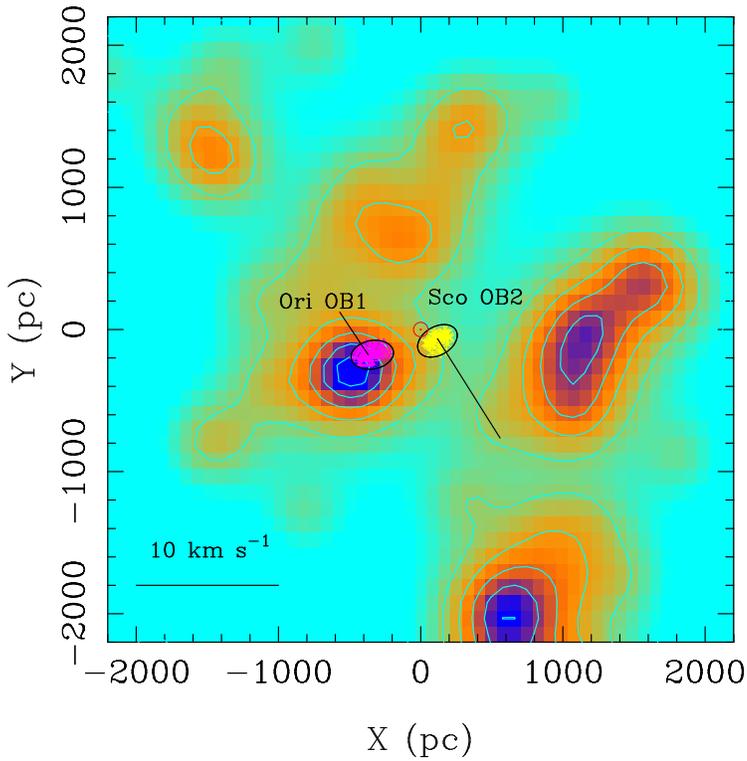}
\caption{Density map for the star clusters younger than 10 Myr,
within a box 4 kpc of side centered in the Sun (red dotted circle).
The Ori OB1 and Sco OB2 associations have been superimposed on the
map, along with their respective residual velocity vectors (black
lines).}
\end{center}
\end{figure}
\clearpage

\clearpage
\begin{figure}
\begin{center}
\includegraphics[angle=270,scale=0.60]{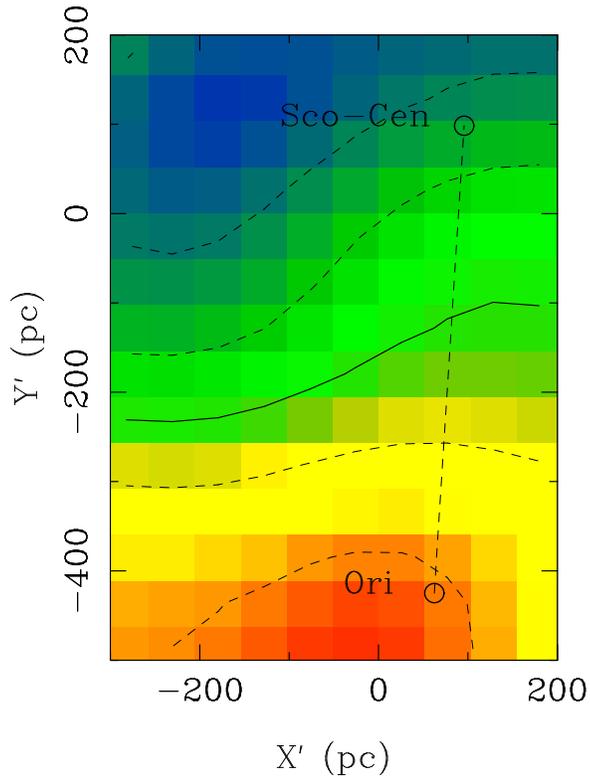}
\caption{Density distribution of the Cluster Formation Index (CFI)
in the GB plane. The open circles represent the centroids of the Ori
OB1 and Sco OB2 associations. The solid contour line represents the
CFI value 0.5, and the dashed contour lines are separated by a value
of the CFI of 0.05}
\end{center}
\end{figure}
\clearpage

\clearpage
\begin{figure}
\begin{center}
\includegraphics[angle=270,scale=0.40]{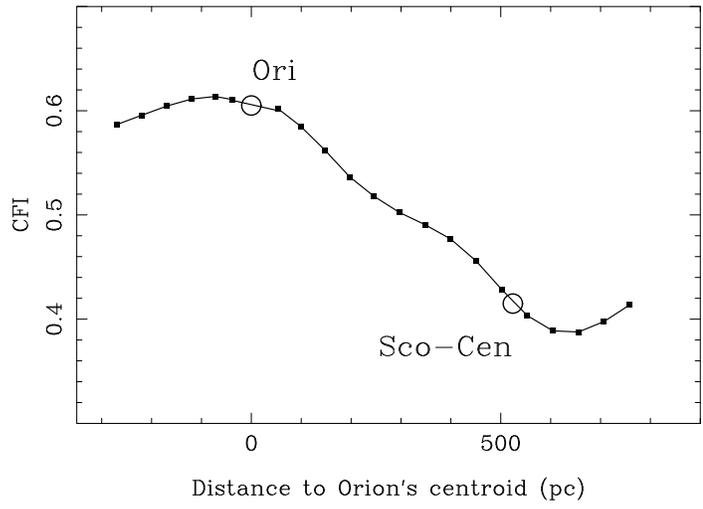}
\caption{Value of the CFI along the line that joins the Ori OB1 and
Sco OB2 centroids (from Figure 10)}
\end{center}
\end{figure}
\clearpage

\clearpage
\begin{figure}
\begin{center}
\includegraphics[angle=270,scale=0.60]{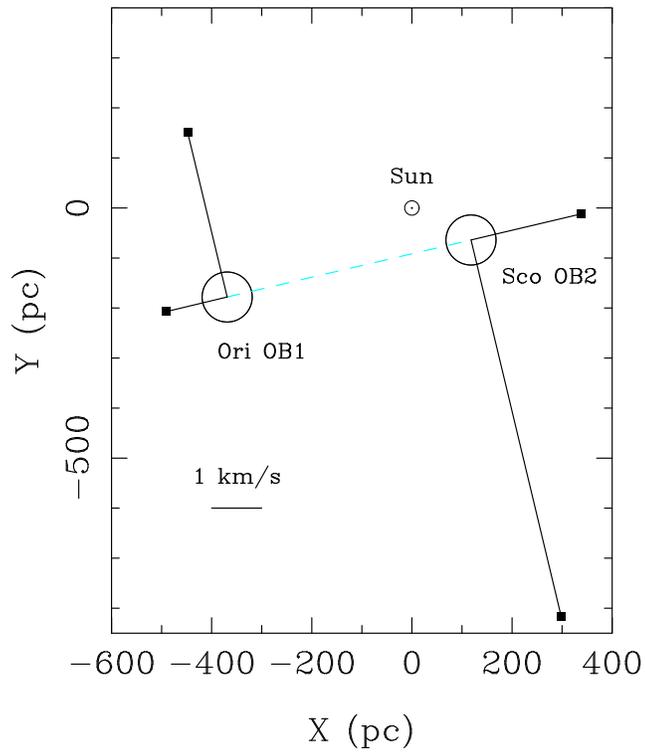}
\caption{``Reduced GB", showing the system as a schematic
composition of the centroids of Ori OB1 and Sco OB2 and their
residual velocities (listed in table~\ref{tbl-2})}
\end{center}
\end{figure}

\begin{table*}
\centering \caption{Catalogue of stellar clusters used in this
study. ($X$,$Y$,$Z$) are the spatial coordinates of the clusters in
the directions of Galactic center, Galactic rotation and North
Galactic pole. ($U$,$V$,$W$) are their respective heliocentric space
velocities, and ($U'$,$V'$,$W'$) their residual velocities corrected
from solar motion and Galactic differential rotation. R$_{cl}$ and
R$_{co}$ stand for cluster radius and core radius, respectively. $T$
is the cluster age. $P$ is the GB membership probability calculated
with our method exclusively by spatial criteria. $Class$ describes
whether a cluster is probably bound (1) or unbound (2).
\label{tbl-1}}
\begin{sideways}
\begin{tabular}{rlrrrrrrrrrrrrrc}
\hline

COCD & Name & $X$ & $Y$ &  $Z$ &  $U$ &  $V$ &  $W$ &  $U'$ &  $V'$ &  $W'$ &  R$_cl$ &  R$_co$ &  $log(T)$ &  $P$ &  $Class$\\
  &   &  ($pc$) &  ($pc$) & ($pc$) &  ($km s^{-1}$) &  ($km s^{-1}$) &  ($km s^{-1}$) &  ($km s^{-1}$) &  ($km s^{-1}$) &  ($km s^{-1}$) &  ($pc$) &  ($pc$) &  ($yr$) &
 (\%) &  \\

\hline
$2$ &  Berkeley 59          & $-471$ & $878$ & $87$ & $20.8$ & $3.1$ & $6.8$ &$2.0$ & $15.7$ & $13.1$ & $3.0$ & $1.0$& $6.80$ & $10$ & $2$\\
$32$ &  Stock 7             & $-491$ & $496$ & $1$ & $10.5$ & $9.7$ & $-7.9$ &$3.9$ & $22.3$&-$1.6$ & $2.2$ & $0.9$& $7.13$ & $7$ & $1$\\
$35$ &  Trumpler 2          & $-476$ & $438$ & $-45$ & $22.1$ & $-34.9$ & $-11.3$ &$17.3$&-$22.3$&-$5.0$ & $5.7$ & $1.7$& $7.93$ & $54$ & $1$\\
$41$ &  Stock 23            & $-291$ & $244$ & $14$ & $13.4$ & $-9.9$ & $-3.6$ &$14.9$ & $2.7$ & $2.7$ & $1.2$ & $0.5$& $7.51$ & $17$& $1$\\
$42$ &  Melotte 20          & $-159$ & $102$ & $-21$ & $-12.9$ & $-26.4$ & $-7.5$ &-$6.8$&-$13.8$&-$1.2$ & $20.8$ & $4.3$& $7.55$ & $50$ & $2$\\
$46$ &  IC 348              & $-354$ & $125$ & $-121$ & $-16.9$ & $-11.6$ & $-8.3$ &-$11.6$ & $1.0$&-$2.0$ & $0.9$ & $0.4$& $7.79$ & $94$ & $1$\\
$68$ &  Collinder 65        & $-301$ & $-47$ & $-58$ & $-14.8$ & $-11.3$ & $-6.9$ &-$4.0$ & $1.3$&-$0.6$ & $13.0$ & $5.7$& $7.41$ & $67$ & $2$\\
$72$ &  Collinder 69        & $-413$ & $-113$ & $-92$ & $-28.3$ & $-11.7$ & $-7.9$ &-$15.4$ & $0.9$&-$1.6$ & $4.2$ & $1.4$& $6.76$ & $84$ & $2$\\
$73$ &  NGC 1981            & $-334$ & $-178$ & $-130$ & $-24.6$ & $-11.9$ & $-6.5$ &-$9.6$ & $0.7$&-$0.2$ & $1.7$ & $0.9$& $7.50$ & $95$ & $1$\\
$74$ &  NGC 1976            & $-329$ & $-183$ & $-132$ & $-23.2$ & $-16.3$ & $-7.1$ &-$8.1$&-$3.7$&-$0.8$ & $2.8$ & $1.4$& $7.71$ & $95$ & $1$\\
$75$ &  NGC 1977            & $-415$ & $-225$ & $-164$ & $-18.7$ & $-17.7$ & $-5.4$ &-$2.2$&-$5.1$ & $0.9$ & $1.8$ & $0.8$& $7.08$ & $97$ & $1$\\
$76$ &  NGC 1980            & $-451$ & $-255$ & $-184$ & $-20.3$ & $-13.6$ & $-7.1$ &-$2.9$&-$1.0$&-$0.8$ & $2.0$ & $1.2$& $6.67$ & $98$ & $2$\\
$77$ &  Collinder 70        & $-338$ & $-158$ & $-117$ & $-16.3$ & $-9.2$ & $-5.8$ &-$1.9$ & $3.4$ & $0.5$ & $10.2$ & $2.7$& $6.71$ & $95$ & $2$\\
$80$ &  Sigma Ori           & $-340$ & $-172$ & $-119$ & $-25.1$ & $-16.4$ & $-3.5$ &-$10.3$&-$3.8$ & $2.8$ & $2.8$ & $0.7$& $6.82$ & $95$ & $2$\\
$91$ &  Platais 6           & $-313$ & $-148$ & $-38$ & $-21.2$ & $-13.1$ & $-12.5$ &-$7.2$&-$0.5$&-$6.2$ & $9.5$ & $3.3$& $7.79$ & $27$ & $2$\\
$93$ &  Collinder 89        & $-789$ & $-120$ & $53$ & $-24.8$ & $-11.8$ & $-0.7$ &-$11.7$ & $0.8$ & $5.6$ & $7.0$ & $2.1$& $7.50$ & $0$ & $1$\\
$95$ &  NGC 2232            & $-265$ & $-183$ & $-42$ & $-12.6$ & $-9.0$ & $-9.7$ &$2.5$ & $3.6$&-$3.4$ & $3.4$ & $0.7$& $7.49$ & $38$ & $1$\\
$107$ &  NGC 2264           & $-607$ & $-257$ & $25$ & $-14.4$ & $-12.5$ & $-11.8$ &$3.1$ & $0.1$&-$5.5$ & $8.4$ & $1.8$& $6.81$ & $1$ & $2$\\
$125$ &  Alessi 21          & $-363$ & $-344$ & $0$ & $-35.3$ & $-19.1$ & $-4.3$ &-$15.0$&-$6.5$ & $2.0$ & $4.4$ & $0.9$& $7.47$ & $6$ & $1$\\
$126$ &  Collinder 132      & $-183$ & $-362$ & $-66$ & $-24.5$ & $-17.5$ & $-10.4$ &-$3.6$&-$4.9$&-$4.1$ & $2.2$ & $0.7$& $7.51$ & $74$ & $1$\\
$133$ &  Collinder 135      & $-112$ & $-292$ & $-62$ & $-17.9$ & $-7.0$ & $-13.9$ &$0.7$ & $5.6$&-$7.6$ & $6.1$ & $1.1$& $7.54$ & $82$ & $1$\\
$136$ &  Collinder 140      & $-168$ & $-361$ & $-55$ & $-21.3$ & $-13.4$ & $-14.3$ &-$0.4$&-$0.8$&-$8.0$ & $3.5$ & $1.4$& $7.57$ & $62$ & $1$\\
$143$ &  Bochum 4           & $-528$ & $-694$ & $12$ & $-13.8$ & $-9.7$ & $-4.9$ &$17.7$ & $2.9$ & $1.4$ & $2.0$ & $0.9$& $7.25$ & $1$ & $1$\\
$155$ &  Haffner 13         & $-301$ & $-646$ & $-46$ & $-52.5$ & $-49.7$ & $-14.4$ &-$22.6$&-$37.1$&-$8.1$ & $6.0$ & $1.2$& $7.51$ & $12$ & $1$\\
$159$ &  NGC 2451A          & $-56$ & $-178$ & $-25$ & $-26.2$ & $-12.8$ & $-13.4$ &-$11.2$&-$0.2$&-$7.1$ & $5.4$ & $2.5$& $7.76$ & $53$ & $1$\\
$162$ &  NGC 2451B          & $-132$ & $-406$ & $-50$ & $-20.4$ & $-6.0$ & $-15.3$ &$1.9$ & $6.6$&-$9.0$ & $3.0$ & $1.5$& $7.88$ & $58$ & $1$\\
$182$ &  Vel OB2            & $-48$ & $-404$ & $-57$ & $-22.6$ & $-12.6$ & $-3.0$ &-$0.3$ & $0.0$ & $3.3$ & $8.6$ & $3.6$& $7.26$ & $78$ & $2$\\
$183$ &  NGC 2547           & $-44$ & $-450$ & $-68$ & $-18.9$ & $-11.1$ & $-13.8$ &$4.8$ & $1.5$&-$7.5$ & $2.4$ & $1.0$& $7.70$ & $78$ & $1$\\
$186$ &  NGC 2546           & $-240$ & $-886$ & $-33$ & $-37.4$ & $-26.8$ & $-9.3$ &$0.2$&-$14.2$&-$3.0$ & $9.8$ & $4.0$& $7.92$ & $7$ & $2$\\
$190$ &  vdBergh-Hagen 23   & $-120$ & $-420$ & $-8$ & $-24.8$ & $-10.9$ & $-5.0$ &-$2.1$ & $1.7$ & $1.3$ & $2.4$ & $0.8$& $7.14$ & $25$ & $1$\\
$202$ &  IC 2391            & $1$ & $-175$ & $-21$ & $-27.7$ & $-15.6$ & $-6.1$ &-$12.8$&-$3.0$ & $0.2$ & $3.8$ & $1.1$& $7.88$ & $57$ & $1$\\
$204$ &  Mamajek 1          & $37$ & $-90$ & $-39$ & $-13.5$ & $-18.6$ & $-10.6$ &-$1.3$&-$6.0$&-$4.3$ & $0.7$ & $0.2$& $6.90$ & $57$ & $2$\\
$205$ &  IC 2395            & $-42$ & $-706$ & $-44$ & $-16.8$ & $-23.6$ & $-8.1$ &$15.1$&-$11.0$&-$1.8$ & $4.3$ & $1.2$& $7.08$ & $41$ & $1$\\
$210$ &  Trumpler 10        & $-53$ & $-414$ & $5$ & $-25.8$ & $-12.0$ & $-10.6$ &-$3.3$ & $0.6$&-$4.3$ & $5.5$ & $1.5$& $7.38$ & $35$ & $1$\\
$213$ &  vdBergh-Hagen 56   & $-65$ & $-677$ & $17$ & $-23.3$ & $20.3$ & $0.3$ &$7.7$ & $32.9$ & $6.6$ & $3.6$ & $1.1$& $7.24$ & $26$ & $1$\\
$216$ &  Platais 8          & $20$ & $-147$ & $-20$ & $-13.0$ & $-18.8$ & $-3.7$ &$1.0$&-$6.2$ & $2.6$ & $7.9$ & $1.6$& $7.75$ & $62$ & $1$\\
$255$ &  vdBergh-Hagen 99   & $151$ & $-509$ & $-5$ & $-28.7$ & $-14.6$ & $-16.2$ &-$3.1$&-$2.0$&-$9.9$ & $3.0$ & $1.2$& $7.86$ & $68$ & $1$\\
$259$ &  IC 2602            & $53$ & $-150$ & $-14$ & $-8.2$ & $-23.1$ & $-0.4$ &$5.9$&-$10.5$ & $5.9$ & $7.0$ & $1.4$& $7.83$ & $62$ & $1$\\
$261$ &  Alessi 5           & $123$ & $-378$ & $-14$ & $-22.3$ & $-20.5$ & $-7.2$ &-$0.9$&-$7.9$&-$0.9$ & $2.1$ & $0.7$& $7.71$ & $66$ & $1$\\
$357$ &  vdBergh-Hagen 164  & $382$ & $-392$ & $-58$ & $-13.4$ & $-29.2$ & $-19.1$ &$8.5$&-$16.6$&-$12.8$ & $3.8$ & $1.3$& $7.14$ & $13$ & $1$\\
$366$ &  NGC 6025           & $623$ & $-444$ & $-79$ & $-12.5$ & $-12.2$ & $3.4$ &$11.0$ & $0.4$ & $9.7$ & $5.1$ & $1.3$& $7.96$ & $3$ & $1$\\
$371$ &  NGC 6087           & $758$ & $-479$ & $-85$ & $-13.9$ & $-4.7$ & $-1.9$ &$10.7$ & $7.9$ & $4.4$ & $7.5$ & $2.2$& $7.93$ & $1$ & $1$\\
$395$ &  NGC 6322           & $961$ & $-252$ & $-53$ & $-56.4$ & $11.3$ & $-6.8$ &-$39.1$ & $23.9$&-$0.5$ & $2.1$ & $1.0$& $7.16$ & $0$ & $1$\\
$402$ &  NGC 6383           & $982$ & $-74$ & $1$ & $3.6$ & $-2.2$ & $-10.5$ &$15.3$ & $10.4$&-$4.2$ & $4.3$ & $1.4$& $6.71$ & $0$ & $2$\\
\hline
\end{tabular}
\end{sideways}
\end{table*}
\begin{table*}
\begin{sideways}
\begin{tabular}{rlrrrrrrrrrrrrrc}
\hline

COCD & Name & $X$ & $Y$ &  $Z$ &  $U$ &  $V$ &  $W$ &  $U'$ &  $V'$ &  $W'$ &  R$_cl$ &  R$_co$ &  $log(T)$ &  $P$ &  $Class$\\
  &   &  ($pc$) &  ($pc$) & ($pc$) &  ($km s^{-1}$) &  ($km s^{-1}$) &  ($km s^{-1}$) &  ($km s^{-1}$) &  ($km s^{-1}$) &  ($km s^{-1}$) &  ($pc$) &  ($pc$) &  ($yr$) &
 (\%) &  \\

\hline
$408$ &  NGC 6405           & $486$ & $-29$ & $-6$ & $-12.3$ & $-12.1$ & $-4.4$ &-$2.1$ & $0.5$ & $1.9$ & $3.4$ & $0.9$& $7.91$ & $6$ & $1$\\
$412$ &  IC 4665            & $290$ & $171$ & $103$ & $-1.4$ & $-14.2$ & $-7.5$ &$2.4$&-$1.6$&-$1.2$ & $6.1$ & $1.8$& $7.63$ & $92$ & $1$\\
$423$ &  Collinder 359      & $542$ & $310$ & $140$ & $10.4$ & $-21.4$ & $-13.6$ &$9.8$&-$8.8$&-$7.3$ & $12.3$ & $4.5$& $7.45$ & $99$ & $2$\\
$425$ &  NGC 6514           & $810$ & $101$ & $-4$ & $-2.4$ & $1.3$ & $-10.8$ &$3.7$ & $13.9$&-$4.5$ & $3.3$ & $1.7$& $7.28$ & $0$ & $1$\\
$449$ &  IC 4725            & $601$ & $146$ & $-48$ & $2.2$ & $-16.6$ & $0.6$ &$6.8$&-$4.0$ & $6.9$ & $6.5$ & $2.2$& $7.83$ & $2$ & $1$\\
$456$ &  Stephenson 1       & $142$ & $331$ & $99$ & $-4.5$ & $-19.4$ & $-10.2$ &-$5.8$&-$6.8$&-$3.9$ & $5.7$ & $1.3$& $7.69$ & $83$ & $1$\\
$479$ &  Roslund 5          & $133$ & $396$ & $2$ & $-5.8$ & $-18.9$ & $-6.9$ &-$9.2$&-$6.3$&-$0.6$ & $3.9$ & $1.1$& $7.77$ & $28$ & $1$\\
$484$ &  Collinder 419      & $153$ & $723$ & $36$ & $23.2$ & $-12.9$ & $-6.7$ &$9.3$&-$0.3$&-$0.4$ & $3.2$ & $0.9$& $6.85$ & $47$ & $2$\\
$500$ &  IC 1396            & $-134$ & $822$ & $54$ & $19.4$ & $-11.9$ & $-7.2$ &$2.4$ & $0.7$&-$0.9$ & $9.5$ & $3.6$& $6.69$ & $56$ & $2$\\
$501$ &  NGC 7160           & $-191$ & $766$ & $89$ & $19.0$ & $-22.4$ & $-1.0$ &$3.8$&-$9.8$ & $5.3$ & $3.2$ & $0.7$& $7.66$ & $43$ & $1$\\
$506$ &  Pismis-Moreno 1    & $-258$ & $858$ & $83$ & $6.4$ & $-22.0$ & $-4.3$ &-$11.7$&-$9.4$ & $2.0$ & $2.5$ & $1.1$& $7.55$ & $34$ & $1$\\
$510$ &  Cep OB3            & $-253$ & $651$ & $36$ & $12.0$ & $-11.3$ & $-4.8$ &$0.4$ & $1.3$ & $1.5$ & $12.2$ & $3.7$& $7.44$ & $31$ & $2$\\
$1013$ & ASCC 13            & $-765$ & $228$ & $45$ & $5.7$ & $-5.3$ & $-6.4$ &$7.7$ & $7.3$&-$0.1$ & $9.8$ & $3.5$& $7.71$ & $1$ & $2$\\
$1016$ & ASCC 16            & $-408$ & $-156$ & $-145$ & $-0.6$ & $-1.5$ & $0.9$ &$13.7$ & $11.1$ & $7.2$ & $5.0$ & $2.0$& $6.93$ & $97$ & $2$\\
$1018$ & ASCC 18            & $-439$ & $-178$ & $-159$ & $-11.9$ & $-6.0$ & $-2.5$ &$3.1$ & $6.6$ & $3.8$ & $5.4$ & $1.7$& $7.12$ & $98$ & $1$\\
$1019$ & ASCC 19            & $-299$ & $-139$ & $-117$ & $-16.0$ & $-9.0$ & $-6.1$ &-$2.3$ & $3.6$ & $0.2$ & $4.9$ & $1.2$& $7.64$ & $94$ & $1$\\
$1020$ & ASCC 20            & $-399$ & $-158$ & $-136$ & $-16.1$ & $-5.3$ & $-5.0$ &-$1.7$ & $7.3$ & $1.3$ & $5.9$ & $2.0$& $7.35$ & $97$ & $1$\\
$1021$ & ASCC 21            & $-451$ & $-163$ & $-142$ & $-17.4$ & $-8.3$ & $-5.3$ &-$2.9$ & $4.3$ & $1.0$ & $7.0$ & $2.6$& $7.11$ & $98$ & $2$\\
$1024$ & ASCC 24            & $-318$ & $-236$ & $-57$ & $-10.2$ & $-10.3$ & $-15.1$ &$6.6$ & $2.3$&-$8.8$ & $2.4$ & $0.8$& $6.96$ & $47$ & $2$\\
$1033$ & ASCC 33            & $-434$ & $-661$ & $-121$ & $-23.6$ & $-1.0$ & $-8.8$ &$6.8$ & $11.6$&-$2.5$ & $12.6$ & $6.3$& $7.26$ & $81$ & $2$\\
$1047$ & ASCC 47            & $-183$ & $-881$ & $5$ & $-41.9$ & $-2.6$ & $-17.6$ & $-4.4$ & $10.0$ & $-11.3$ & $7.9$ & $2.4$ & $7.88$ & $10$ & $1$\\
$1050$ & ASCC 50            & $-111$ & $-843$ & $22$ & $-30.4$ & $-13.5$ & $-9.6$ & $5.9$ & $-0.9$ & $-3.3$ & $5.9$ & $1.8$ & $7.48$ & $17$ & $1$\\
$1058$ & ASCC 58            & $122$ & $-587$ & $14$ & $-33.0$ & $-15.3$ & $-13.0$ & $-4.9$ & $-2.7$ & $-6.7$ & $4.2$ & $1.1$ & $7.04$ & $60$ & $1$\\
$1069$ & ASCC 69            & $482$ & $-867$ & $-126$ & $-29.8$ & $-17.5$ & $-9.4$ & $7.2$ & $-4.9$ & $-3.1$ & $7.0$ & $2.1$ & $7.91$ & $11$ & $1$\\
$1076$ & ASCC 76            & $376$ & $-465$ & $-44$ & $-19.7$ & $2.7$ & $-5.4$ & $4.5$ & $15.3$ & $0.9$ & $3.7$ & $1.1$ & $7.45$ & $15$ & $1$\\
$1079$ & ASCC 79            & $612$ & $-513$ & $-40$ & $-8.1$ & $-15.3$ & $-7.9$ & $17.7$ & $-2.7$ & $-1.6$ & $7.3$ & $3.5$ & $6.86$ & $3$ & $2$\\
$1084$ & ASCC 84            & $721$ & $-532$ & $-85$ & $-16.2$ & $-11.7$ & $-10.7$ & $10.2$ & $0.9$ & $-4.4$ & $3.9$ & $1.3$ & $7.68$ & $1$ & $1$\\
$1104$ & ASCC 104           & $460$ & $654$ & $-22$ & $9.3$ & $-13.7$ & $-8.2$ & $-2.3$ & $-1.1$ & $-1.9$ & $6.7$ & $2.1$ & $7.71$ & $2$ & $1$\\
$1105$ & ASCC 105           & $229$ & $444$ & $19$ & $-2.5$ & $-19.5$ & $-7.0$ & $-7.4$ & $-6.9$ & $-0.7$ & $5.2$ & $1.8$ & $8.00$ & $31$ & $1$\\
$1114$ & ASCC 114           & $-68$ & $546$ & $10$ & $15.5$ & $-20.2$ & $-0.8$ & $7.4$ & $-7.6$ & $5.5$ & $1.5$ & $0.8$ & $7.75$ & $62$ & $1$\\
$1118$ & ASCC 118           & $-224$ & $869$ & $66$ & $27.4$ & $-35.6$ & $-11.6$ & $8.9$ & $-23.0$ & $-5.3$ & $3.3$ & $1.6$ & $7.02$ & $41$ & $1$\\
$1127$ & ASCC 127           & $-132$ & $323$ & $25$ & $-5.5$ & $-10.5$ & $-8.4$ & $-6.6$ & $2.1$ & $-2.1$ & $4.4$ & $1.1$ & $7.82$ & $42$ & $1$\\
\hline
\end{tabular}
\end{sideways}
\end{table*}

\begin{table*}
\caption{Centroids of the OB associations Ori OB1, and Sco OB2, in
phase space\label{tbl-2}}
\begin{tabular}{rrrrrrr}
\hline
Name & $X$ & $Y$ & $Z$ & $U'$ & $V'$ & $W'$\\
 & ($pc$) & ($pc$) & ($pc$) & ($km s^{-1}$) & ($km s^{-1}$) & ($km s^{-1}$)\\
\hline
Ori OB1 &  $-369 \pm 7$ & $-178 \pm 6$ & $-132 \pm 4$ & $-2 \pm 1$ & $3 \pm 1$ & $1 \pm 0.6$\\
Sco OB2 &  $118 \pm 7$  & $-64 \pm 6$  & $29 \pm 4$  & $4 \pm 1$ & $-7 \pm 1$ & $0 \pm 0.5$\\
\hline
\end{tabular}
\end{table*}

\begin{table*}
 \centering
\caption{Comparison of the ``reduced GB" parameters with their range
of values found in the literature\label{tbl-3}}
\begin{tabular}{lcccc}
\hline
 &  $i$ & $\Omega$ &  $\rho_0$ & $\omega_0$\\
 & ($\degr$) & ($\degr$) & ($km s^{-1} kpc^{-1}$) & ($km s^{-1} kpc^{-1}$)\\
\hline
Reduced GB    &  $18$ & $283$ & $7$ & $22$\\
Literature    &  $14-27$ & $271-290$ & $0-29$ & $12-37$\\
\hline
\end{tabular}
\end{table*}

\bsp

\label{lastpage}

\end{document}